\documentstyle [12pt] {article}

\newcommand{\cs}[3]{{{#3} \brace {#1 #2}}}

\begin{document}
\begin{center}
\bf {An Alternative Source for Dark Energy  }\\

\end{center}
\begin{center}
\bf{M.I.Wanas\footnote{Astronomy Department, Faculty of Science,
Cairo University, Giza, Egypt.

E-mail:wanas@frcu.eun.eg}}
\end{center}

\begin{abstract}
In the present work, an alternative interpretation of the source
of accelerated expansion of the Universe is suggested. A probable
candidate is the interaction between the quantum spin of a moving
particle and the torsion of space-time, produced by the background
gravitational field of the Universe. This interaction has been
suggested by the author in a previous work, with some experimental
and observational evidences for its existence. It has been shown
that this interaction gives rise to a repulsive force. The
accelerated expansion of the Universe may give a further evidence
on the existence of this interaction on the cosmological scale.
\end{abstract}
\section*{}

 Recently, observation of red-shift of type $Ia$ supernova [1]
indicate that our Universe is in an accelerating expansion phase.
The four known fundamental interactions, as we understand them
today, can not account for the accelerating expansion of the
universe. Weak and strong interactions can be easily ruled out,
since both are very short range and cannot play any role in the
large scale behavior of our Universe. Also, electromagnetism
cannot play any role, in this respect, since neither the universe
nor its constituents are electrically charged, while monopoles are
not present in our Universe. The only interaction which is playing
the main role, in the structure and evolution of the universe, is
gravity. But gravity, as far as we know, is attractive and not
repulsive. Consequently, it can not give any satisfactory
interpretation of the accelerating phase of the Universe. So, what
is real cause of the accelerating expansion of the universe? Is
the existing physics sufficient to deal with this problem? Do we
need new physics?

Theories of gravity consider the interaction between matter
(energy) and the gravitational field. The General Theory of
Relativity (GR), the most successful gravity theory so far,
considers gravity as a space-time property. Its field equations
shows how matter produces gravity. i.e. how matter affects
space-time structure. Its equation of motion shows how space-time
structure (gravity) affects trajectories of test particles
(matter). In the context of GR, this interaction is between
space-time structure (curvature) and one of properties of matter
(mass or energy). Mass (energy) of a material object is collection
of the masses (energies) of its elementary constituents. These
constituents have other properties beside mass, e.g. spin, charge,
... etc. GR does not answer questions  neither about the
interaction between space-time structure and such other properties
of matter, nor whether such interactions exists or not.

If we consider geometries with curvatures and torsion, in
particular the {\it{Parameterized Absolute
Parallelism}}(PAP)-geometry [2], it has been shown that the
parameterized path equation of this geometry can be written as
[3].
$$
\frac{d^2 x^{\mu}}{d s^2} + \cs{\alpha}{\beta}{\mu}
\frac{dx^{\alpha}}{d s}\frac{dx^{\beta}}{d s} = - b \Lambda^{..~
\mu}_{\alpha \beta.}\frac{dx^{\alpha}}{d s}\frac{dx^{\beta}}{d s}
\eqno(1) $$
where $\cs{\alpha}{\beta}{\mu}$ is the Christoffel
symbol, $\Lambda^{\mu}_{. \alpha \beta}$ is the torsion of the
space-time and $b$ is a dimensionless parameter. The term on the
R.H.S. of this equation has been suggested to represent an
interaction between the spin of the moving particle and the
torsion of space-time. In the presence of this interaction, the
total potential felt by a spinning particle, in the linearization
regime of (1), is found to be [3]
$$
\Phi_{s} = (1-b) \Phi_{N} \eqno(2)
$$
where $\Phi_{N}$ is the Newtonian potential. The second term with
the (-)sign on the R.H.S. of (2)gives rise to a repulsive force.
This force is needed to interpret the results of the observation
of SN type Ia [1].

For some experimental and observational reasons, the parameter $b$
is suggested to the form
$$
b = \frac{n}{2} \alpha \gamma  \eqno(3)
$$
where $n$ is a natural number is, $\alpha$ is the fine structure
constant and $\gamma$ is a dimensionless parameter characterizes
the source of the background field. Equation (2) has been used to
interpret the discrepancy of in the results of the COW-experiment
[4] considering the gravitational field of the Earth as weak,
static field and taking $n=1$ (for neutrons)and $\gamma =1$ for
the Earth. The same equation has been used to construct a temporal
model of SN1987A [5]. Also, it has been used to study the effect
of spin-torsion interaction on the Chandrasekhar-limit [6].

Many authors have tried to interpret the accelerating expansion of
the universe using different assumptions. Most of them consider
the cosmological term in the equations of General Relativity, as a
probable candidate (for a review of alternative interpretations
see [7]). Others consider special types of matter to be the source
of dark energy [8]!

In conclusion we draw the following comments: \\
1. Torsion of space-time should be taken into account in order to
complete our understanding of physical phenomena, especially on
large scale. \\
2. Spin-torsion interaction, giving rise to a repulsive force, can
be used to interpret the accelerating expansion of the Universe,
and may prevent the existence of space-time singularities. \\
3. Torsion may give a reasonable geometric origin for the
cosmological term. \\
4. A field theory, taking into account the torsion of space-time
is needed now in order to find the type of dependence of the
parameter $\gamma$ on the energetic contents of any system under
consideration.

An extended version of the present work will be published
elsewhere. \\
\section*{Acknowledgement}
The author would like to thank the authorities of the ICTP
Trieste, Italy for giving him a grant to participate in this
conference.

\section*{References}
{[1]} Tonry, J.L., Schmdit, B.P. et al.
(2003) Astrophys. J.
{\bf{594}}, 1. \\
{[2]} Wanas, M.I. (2000) Turk. J. Phys. {\bf{24}}, 473; gr-qc/0010099 \\
{[3]} Wanas, M.I. (1998) Astrophys. Space Sci., {\bf{258}}, 237; gr-qc/9904019 \\
{[4]} Wanas, M.I., Melek, M. and Kahil, M.E. (2000) Gravit.
Cosmol.,{\bf{6}}, 319.\\
 {[5]} Wanas, M.I., Melek, M. and Kahil, M.E. (2002) Proc. MG IX
Part B, p.1100 ;

 gr-qc/0306086 \\
{[6]} Wanas, M.I. (2003)Gravit. Cosmol.,{\bf{9}}, 109.\\
{[7]} Mannheim, P.D. (2006) Prog.Part.Nucl.Phys. {\bf{56}} 340; gr-qc/0505266 \\
{[8]} Babourova, O.V. and Frolov, B. (2003) Class.Quant.Grav. {\bf{20}}1423 ;

gr-qc/0209077 \\

\end{document}